# Spin–orbit-coupled metal candidate PbRe$_2$O$_6$

Satoshi Tajima, Daigorou Hirai, Takeshi Yajima, Daisuke Nishio-Hamane, Yasuhito Matsubayashi[a] and Zenji Hiroi[*]

*Institute for Solid State Physics, University of Tokyo, Kashiwa, Chiba 277-8581, Japan*

We study the lead rhenium oxide PbRe$_2$O$_6$ as a candidate spin–orbit-coupled metal (SOCM), which has attracted much attention as a testing ground for studying unconventional Fermi liquid instability associated with a large spin–orbit interaction. The compound comprises a stack of modulated honeycomb lattices made of Re$_{5+}$ (5$d_2$) ions in a centrosymmetric $R$–3$m$ structure at room temperature. Resistivity, magnetic susceptibility, and heat capacity measurements using single crystals reveal two successive first-order phase transitions at $T_{s1}$ = 265 K and $T_{s2}$ = 123 K. At $T_{s1}$, the magnetic susceptibility is enormously reduced and a structural transition to a monoclinic structure takes place, while relatively small changes are observed at $T_{s2}$. Surprisingly, PbRe$_2$O$_6$ bears a close resemblance to another SOCM candidate Cd$_2$Re$_2$O$_7$ despite crucial differences in the crystal structure and probably in the electronic structure, suggesting that PbRe$_2$O$_6$ is an SOCM.

[a]Present address: Advanced Coating Technology Research Center, National Institute of Advanced Industrial Science and Technology, Tsukuba, Ibaraki 305-8564, Japan
[*]Corresponding author

## 1. Introduction

5$d$ transition metal compounds have recently attracted much attention from solid state chemists who are searching for novel phenomena and new states of matter in the field of solid state physics. Instead of the spin and orbital degrees of freedom of an electron, which are often crucial for 3$d$ transition metal compounds, a multipole degree of freedom can appear in a 5$d$ electron system, which is considered to be a spin–orbital composite generated by a large spin–orbit coupling (SOC) for heavy elements. One expects exotic multipole-ordered states with unusual properties in 5$d$ transition metal compounds. However, compared with well-studied 3$d$ compounds, the number of 5$d$ compounds remains limited. We have been searching for new 5$d$ transition metal compounds or digging old ones out of the ground to be studied in detail by recent advanced experimental techniques and through modern understanding of complex phenomena.

### 1.1. Spin–orbit-coupled metal

The spin–orbit-coupled metal (SOCM) to be focused on in this study refers to an itinerant electron system that exhibits a spontaneous spatial inversion symmetry breaking (ISB) owing to a Fermi liquid instability associated with a large SOC [1]. The low-temperature (LT) phase without inversion symmetry is classified into three categories, odd-parity multipolar phase, metallic ferroelectric phase, and gyrotropic phase, depending on the crystal symmetry of the high-temperature (HT) phase. The ISB transition in the SOCM should cause significant changes in the electronic properties because the spin-degenerate Fermi surface (FS) at high temperatures is rendered spin-split by the antisymmetric SOC activated as a result of ISB, as illustrated in Fig. 1; a large antisymmetric SOC can generate a sizable spin splitting in the SOCM. In the spin-split FS, a spin-momentum locking can lead to unconventional electromagnetic properties such as the Edelstein effect, which are fascinating in the field of spintronics [2]. In addition, one expects that parity fluctuations related to the ISB transition induce exotic superconductivity with parity mixing near the quantum critical point [3,4].

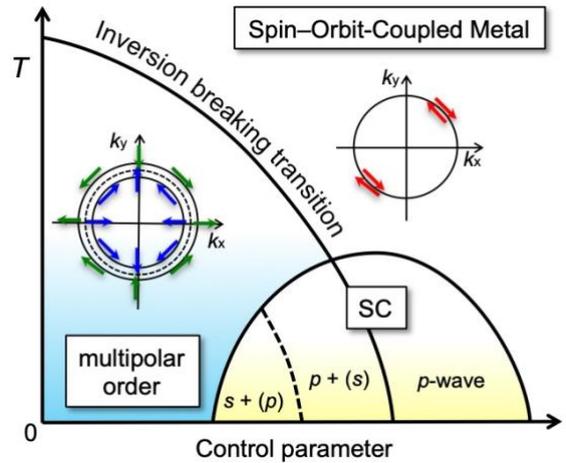

**Fig. 1.** Schematic phase diagram for a spin–orbit-coupled metal (SOCM) [3-5]. The inherent Fermi liquid instability of the SOCM causes a spatial inversion symmetry breaking (ISB) transition and stabilizes an odd-parity multipolar phase at low temperatures, where an antisymmetric spin–orbit coupling is activated, resulting in a spin-split FS as depicted below the ISB line; the blue and green arrows represent the directions of spins varying along the inner and outer FSs, respectively, while the circle with the red arrows above the ISB line represents a spin-degenerate FS. As a function of certain control parameters such as pressure, the ISB line can be suppressed and vanishes at a quantum critical point, near which a superconducting dome composed of three regimes is expected: $s$-wave-dominant region [$s + (p)$], $p$-wave-dominant region [$p + (s)$], and a region with only $p$-wave pairings [4].

The metallic pyrochlore oxide Cd$_2$Re$_2$O$_7$ (CRO) has been considered as a typical candidate SOCM, in which the 5$d_2$ electrons from the Re$_{5+}$ ion are itinerant in the vertex-sharing tetrahedron network [5]. CRO crystallizes in a cubic $Fd$–3$m$ structure with inversion symmetry at room temperature (phase I). An ISB transition takes place at $T_{s1}$ ~ 200 K into the tetragonal phase II of the space group $I$–4$m$2; the threefold rotation axis is simultaneously lost. At this transition, large decreases in the resistivity and density of states (DOS) are observed. In addition, another structural transition to phase III of $I4_122$ occurs at $T_{s2} \sim 120$ K, which is accompanied by relatively small changes in the properties. Then, superconductivity sets in at $T_c = 1.0$ K in the absence of inversion symmetry.

The first ISB transition of CRO has been attributed to the Fermi liquid instability of the SOCM, and the reduction of the DOS is considered to be due to the loss of half of the hole FSs; the spin-split FSs have been experimentally observed by quantum oscillation measurements [6]. Theoretically proposed for the two LT phases are odd-parity multipolar orders, such as the electric toroidal quadrupoles [2,7] or other complex multipoles [8,9]. Hence, the previous studies have clearly demonstrated that CRO is a good candidate SOCM. On the other hand, the superconductivity has been revealed to be of the conventional $s$-wave type, which may be because, at ambient pressure, the compound is located near the left edge of the superconducting dome in the phase diagram of Fig. 1. In fact, signatures for an unconventional superconductivity have been observed under high pressures around 4 GPa, where the ISB line tends to disappear [10].

Another compound thus far proposed to be an SOCM candidate is $LiOsO_3$ with $Os^{5+}$ ($5d^3$) ions. It shows metallic conductivity and an ISB transition from the HT $R\text{–}3c$ structure with inversion symmetry to a polar $R3c$ structure at 140 K [11]. This structural sequence is identical to that of the ferroelectric transition of its insulating analogue $LiNbO_3$. The LT phase has been considered as a "ferroelectric metal" as expected for an SOCM [11,12]. However, the accompanying changes in the electronic properties are relatively small compared with those of CRO. Moreover, the origin of the structural transition has been ascribed to a purely structural instability coming from the mismatch in the ionic radius between Li and the other ions, as in the case of $LiNbO_3$ [13]. Thus, it is unclear whether $LiOsO_3$ really is an SOCM. Therefore, the variety of materials remains limited thus far. To establish the concept of the SOCM in general, more candidate compounds are required for investigations, as the nature of ISB and the properties of the LT phases should depend on various aspects, such as the crystal and electronic structures of the HT phases and the types of spin-splitting in the LT phases.

*1.2. $PbRe_2O_6$*

The present compound $PbRe_2O_6$ (PRO) was prepared by Wentzell et al. in 1985 [14], but has not been studied since the first report to our knowledge. It comprises $Re^{5+}$ ions with the $5d^2$ electron configuration and crystallizes in a rhombohedral structure of space group $R\text{–}3m$ with inversion symmetry; the possibility of the noncentrosymmetric space group $R3m$ was not excluded in the previous structural analysis. Thus, PRO is a candidate SOCM, and we take a great deal of interest in its LT properties. In this study, we prepared single crystals of PRO and measured the bulk properties and examined the crystal structure at low temperatures down to 2 K. It is revealed that the compound is a Pauli paramagnetic metal and exhibits at least two phase transitions at $T_{s1} = 265$ K and $T_{s2} = 123$ K. Below $T_{s1}$, the magnetic susceptibility shows an enormously large decrease and the crystal symmetry is lowered to monoclinic, while smaller changes occur at $T_{s2}$.

These features surprisingly resemble those of CRO, suggesting that PRO is another candidate SOCM.

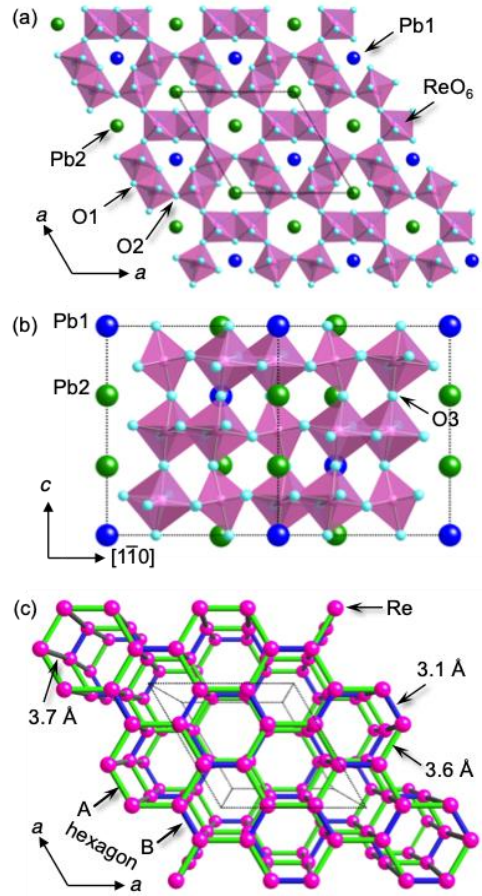

**Fig. 2.** Crystal structure of $PbRe_2O_6$ ($R\text{–}3m$, $a = 10.359$ Å, $c = 11.092$ Å) [14] viewed along the $c$ axis (a) and along the [1$\bar{1}$0] direction perpendicular to the $c$ axis (b). The unit cell is depicted by the dotted lines in each figure. The $ReO_6$ units are illustrated by the magenta octahedra with the sky blue balls representing the oxygen atoms at the vertices. The Pb1 and Pb2 atoms are shown by blue and green balls, respectively. (c) Plane view of the Re sublattice, which forms modulated honeycomb lattices stacked along the $c$ axis. The colored sticks between neighboring Re atoms distinguish three kinds of bonds with lengths of 3.1 and 3.6 Å in the plane and 3.7 Å along the $c$ axis. There are two types of Re hexagons in the plane: regular hexagon A with 3.6 Å bonds and hexagon B with alternating bonds of 3.6 and 3.1 Å, which are stacked in a sequence of ABB along the $c$ axis.

Figure 2 shows the characteristics of the crystal structure of $PbRe_2O_6$ at room temperature [14]. All the Re atoms are located in one crystallographically equivalent site (18$g$ site in $R\text{–}3m$), and each of them is coordinated octahedrally by six oxide ions. The $ReO_6$ octahedra are connected with each other by their vertices and edges to form a layer perpendicular to the $c$ axis, as depicted in Fig. 2a. The layer contains two types of "holes": the hexagonal hole is surrounded by six octahedra connected by vertices and the triangular hole is surrounded by six octahedra by both vertices and edges. Along the $c$ axis, the octahedra are connected via vertices to form a 3D network (Fig. 2b).

On the other hand, when only the Re atoms are highlighted in the structure, a stack of honeycomb lattices appear, as depicted in Fig. 2c. The honeycomb lattice consists of two types of Re hexagons A and B: regular hexagon A with 3.6 Å bonds is located around the hexagonal hole in Fig. 2a, while modulated hexagon B with alternating bonds of 3.6 and 3.1 Å lies around the triangular hole. The stacking sequence of the Re hexagons along the $c$ axis is ABB with a separation of 3.7 Å. Alternatively, when one considers the pair of edge-sharing octahedra or the pair of Re atoms as a building unit, the sublattice could be a stacking of kagome lattices.

There are two kinds of Pb atoms between the Re–O layers: the Pb1 atom at the $3c$ site is sandwiched by two triangular holes, which are inverted by inversion symmetry at the Pb1 atom, and the Pb2 atom at the $6c$ site is located between triangular and hexagonal holes. Note that the Re atom has no inversion symmetry and only the Pb1 atom has it, which is in contrast to the case of CRO having inversion symmetry at the Re atom in the HT phase I. It is pointed out that the inversion symmetry at the Pb1 site can be removed by a deformation of the surrounding Re–O network.

## 2. Experimental Procedure

### 2.1. Sample preparation

Single crystals of PRO were prepared via a solid-state reaction between β-PbO (RARE METALLIC CO. LTD. 99.999%) and $ReO_3$; a powder of $ReO_3$ was prepared in advance by a solid-state reaction of $Re_2O_7$ and Re in a molar ratio of 3:1 in a sealed quartz ampoule at 573 K for 12 h. The two starting ingredients were thoroughly mixed in molar ratios of 1.4–2.6:2 and pelletized in a glove box filled with dry argon gas. The pellet was sealed in a quartz ampoule and heated at 843 K for 24–600 h. The reaction ended to produce many small crystals on the wall of the quartz ampoule near the pellet, as shown in Fig. 3a. Several "large" PRO crystals, such as those shown in Fig. 3b, were obtained after removing the impurity phases of $Pb_6Re_6O_{19}$ and $Pb(ReO_4)_2$ by dissolving them in hydrochloric acid. The thus-prepared crystal possesses a copperish color with metallic luster and a hexagonal prismatic shape of ~1 mm length at most along the $c$ axis and sub-millimeter diameter. Note that sizable crystals were obtained only when the starting mixture had been ground with force in an agate mortar, which might have transformed β-PbO to α-PbO by a mechanochemical reaction [15]. The crystal was gradually degraded in air ambient to tarnish in a week. The chemical composition examined by energy-dispersive X-ray spectroscopy in a scanning electron microscope (JEOL JEM IT-100) yielded Pb:Re = 1.00(2):2.00(2), which is in good agreement with the ideal composition of PRO.

### 2.2. X-ray diffraction measurements

Powder X-ray diffraction (XRD) experiments at room temperature were carried out on crushed crystals in a standard diffractometer with Cu-Kα radiation (Rigaku SmartLab). As shown in Fig. 4, all the diffraction peaks, except those at around 25º from a small amount of impurity of $Pb(ReO_4)_2$, are indexed to a hexagonal cell of $a$ = 10.37013(6) Å and $c$ = 11.11835(9) Å, which are close to the reported values: $a$ = 10.359(6) Å and $c$ = 11.092(6) Å [14]. Thus, in combination with the result of chemical analysis, we conclude that PRO crystals have been successfully prepared.

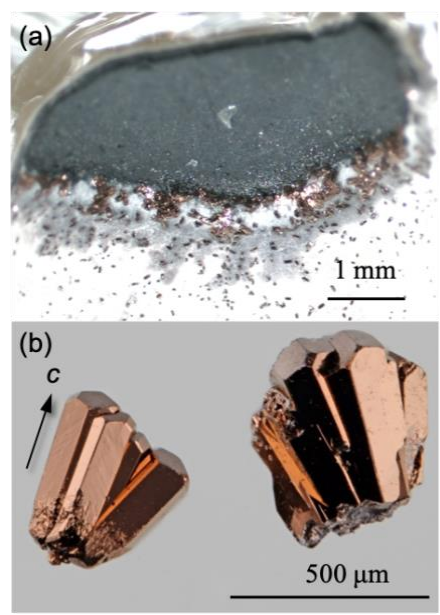

**Fig. 3.** Photographs of a pellet after reaction (a) and typical crystals of PRO (b). Many copperish crystals grow on the inner surface of the quartz tube, particularly large ones of sub-millimeter size such as those shown in (b) near the edge of the pellet.

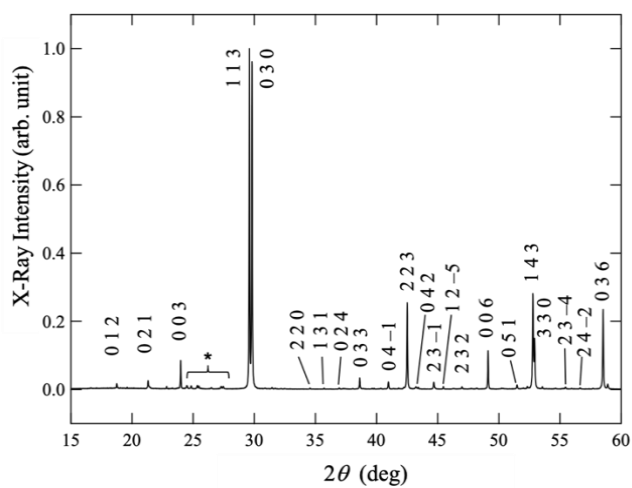

**Fig. 4.** Powder X-ray diffraction profile using standard Cu–Kα radiation from crushed crystals of PRO at 300 K. The indices of the diffraction peaks are based on a hexagonal unit cell of $a$ = 10.37013(6) Å and $c$ = 11.11835(9) Å. The several peaks at around 25º are from an impurity of $Pb(ReO_4)_2$.

Powder XRD experiments at low temperatures were performed in a SmartLab (Rigaku). The incident X-ray beam was monochromated by a Johansson-type monochromator with a Ge(111) crystal to select only Cu–Kα1 radiation, which gave a much higher resolution than in the conventional experiment and made it possible to detect a peak splitting or a broadening associated with even a small symmetry change in the crystal structure. A powder sample from crushed crystals was put on a copper plate that was cooled by a Gifford–McMahon refrigerator down to 4 K. The lattice constants were determined by the whole pattern fitting at 4 K.

### 2.3. Physical property measurements

Electrical resistivity $\rho$ was measured upon cooling and then heating between 2 and 400 K at a rate of 2 K/min by the standard four-probe method in a Quantum Design Physical Property Measurement System (PPMS). Magnetic susceptibility $\chi$ was measured in a Quantum Design Magnetic Property Measurement System 3. Approximately 30 crystals of 3.57 mg weight in total were attached to a semicylindrical rod of quartz with varnish so that their $c$ axes were approximately parallel to each other. Two measurements were carried out with a magnetic field of 7 T parallel and perpendicular to the $c$ axis. The temperature was scanned upon cooling and heating between 2 and 400 K at a rate of 1.5 K/min. The data were collected using the DC mode with a 30 mm scan in 1 s during the temperature scan.

Heat capacity was measured at 2–300 K in a PPMS by the relaxation method. For the measurements, one crystal with 9.58 mg weight was put on a sapphire plate with a small amount of Apiezon grease. The heat capacity of the sample was obtained by fitting a thermal relaxation curve after a heat pulse giving a temperature rise of 2% of the base temperature; an addendum contribution was measured in advance without a sample and subtracted from the data. Note that the latent heat of a first-order transition cannot be evaluated correctly by this relaxation method; a first-order transition generally appears as a sharp peak instead of a jump for a second-order transition.

## 3. Results and discussion

### 3.1. Successive transitions

The electrical resistivity $\rho$ of PbRe$_2$O$_6$ shows a metallic temperature dependence below 400 K, as shown in Fig. 5a. The $\rho$ values are 12 and 0.64 mΩ cm at 300 and 2 K, respectively; the residual resistivity ratio (RRR) is 17. Upon cooling, $\rho$ exhibits two anomalies: a sudden jump at ~265 K and a broad hump at around 120 K. There are small ($\Delta T \sim 1$ K) and large ($\Delta T \sim 20$ K) thermal hystereses, respectively, indicating that both transitions are of the first order. The transition temperatures are defined as the average of the peak temperatures in the temperature derivatives for the cooling and heating curves: $T_{s1} = 265$ K and $T_{s2} = 123$ K. We have observed two similar anomalies for other crystals with a range of RRR values between 4 and 37, confirming that they are intrinsic transitions. The three phases separated by the transitions are named phases I, II, and III in order of decreasing temperature. In our preliminary resistivity measurements, no superconducting transition was observed above 0.5 K.

Figure 5b shows magnetic susceptibilities measured at magnetic fields of 7 T applied parallel ($\chi_c$) and perpendicular ($\chi_{ab}$) to the [001] direction. The weak temperature dependence at 300–400 K indicates a Pauli paramagnetism for an itinerant electron system with a weak electron correlation without a magnetic instability. Upon cooling, each $\chi$ exhibits a sudden drop at $T_{s1} = 265$ K followed by a gradual decrease below $T_{s1}$: the magnitudes of the drop at 265 K correspond to 40% and 30% of the corresponding 300 K values and the minimum values at ~150 K are 30% and 42% for $\chi_c$ and $\chi_{ab}$, respectively. These large reductions in $\chi$ at $T_{s1}$ must be partly due to a significant loss of the DOS, just as in the case of Cd$_2$Re$_2$O$_7$ [5]. Small thermal hystereses are observed at the $T_{s1}$ transition, corresponding to that in $\rho$. Then, the $\chi$ data show small anomalies at around $T_{s2} = 123$ K with large thermal hystereses, again consistent with the transition in $\rho$. Thus, large and small changes in the electronic states should occur at the $T_{s1}$ and $T_{s2}$ transitions, respectively. In addition, we note that there is another small anomaly at ~250 K.

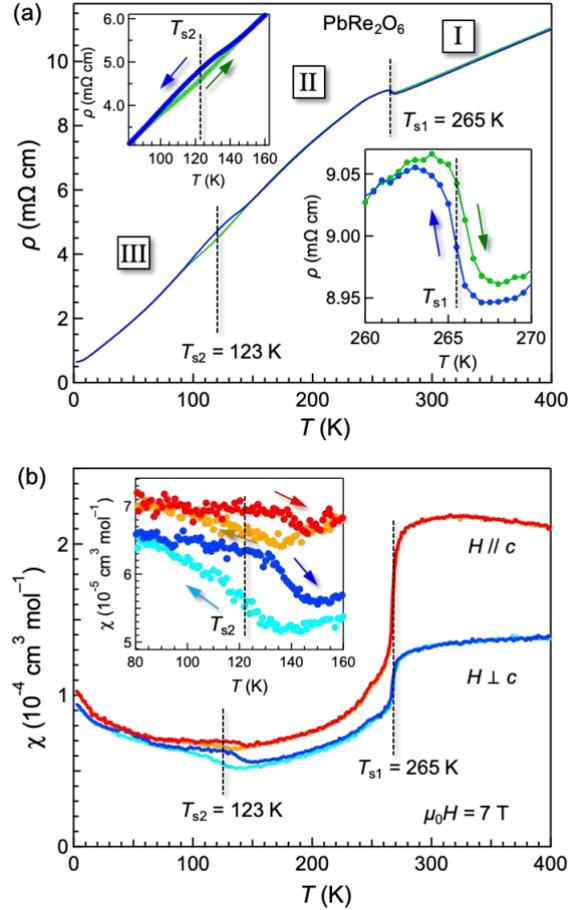

**Fig. 5.** Temperature dependences of resistivity $\rho$ (a) and magnetic susceptibility $\chi$ (b) of PbRe$_2$O$_6$ measured upon cooling and heating. The resistivity was measured on one crystal with the electrical current along the $c$ axis. The magnetic susceptibility was measured on approximately 30 oriented crystals with magnetic fields of 7 T applied parallel ($H // c$) and perpendicular ($H \perp c$) to the $c$ axis of the HT hexagonal structure. Two successive first-order transitions with thermal hystereses are observed at $T_{s1} = 265$ K and $T_{s2} = 123$ K. The insets expand the variations near the transitions. The phases separated by these transitions are called phases I, II, and III in order of decreasing temperature.

Note that there is a large anisotropy in $\chi$ above $T_{s1}$: $\chi$ is $2.17 \times 10^{-4}$ and $1.32 \times 10^{-4}$ cm$^3$ mol$^{-1}$ at 300 K for $\chi_c$ and $\chi_{ab}$, respectively. In general, there are three contributions involved in the $\chi$ data: Pauli paramagnetism or spin susceptibility $\chi_s$ from conduction electrons, Van Vleck paramagnetism or orbital susceptibility $\chi_{orb}$, and core diamagnetism from inner shells, the last of which is isotropic and calculated to be $-1.42 \times 10^{-4}$ cm$^3$ mol$^{-1}$ from the literature [16]. The second term can be large for a heavy transition metal with a large SOC as for Re; in fact, a large $\chi_{orb}$ of $3.16 \times 10^{-4}$ cm$^3$ mol$^{-1}$ was estimated for CRO by Cd NMR experiments, which is nearly equal to $\chi_s$ [17]. For PRO, we expect a similarly large $\chi_{orb}$ although it has not yet been evaluated. Moreover, one expects a certain anisotropy in $\chi_{orb}$ arising from the hexagonal crystal structure. It is likely that a structural transition at $T_{s1}$ mentioned later has removed the orbital degeneracy present in phase I so that $\chi_{orb}$ is significantly reduced in phase II with its anisotropy also being reduced by the transition. Nevertheless, we think that the observed large decrease in $\chi$ at and below $T_{s1}$ also includes a reduction in $\chi_s$ owing to a partial loss in the DOS. Further quantitative discussion requires microscopic information from NMR or photoemission spectroscopy measurements.

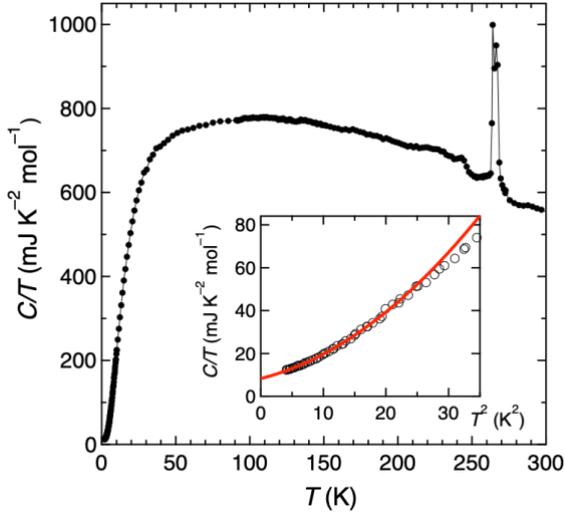

**Fig. 6.** Heat capacity divided by temperature $C/T$ showing sharp spikes at around 265 K for the $T_{s1}$ transition. A peak corresponding to the $T_{s2}$ transition is not discernible, which may be too small and broad to distinguish from the background phonon contribution. There is also a small anomaly at 243 K. The inset shows a $C/T$ vs $T^2$ plot for the low-temperature data below 6 K. A fit of the data at 2–5 K to the equation in the main text, the red curve, gives a Sommerfeld coefficient of 8.3(4) mJ K$^{-2}$ mol$^{-1}$ and a Debye temperature of 288 K.

The $T_{s1}$ transition has been clearly evidenced by the heat capacity data shown in Fig. 6. There are two sharp spikes at 264 and 266 K, giving $T_{s1}$ = 265 K on average, consistent with resistivity and magnetic susceptibility data; the two spikes may be due to an artifact in measuring a first-order transition by the relaxation method. In contrast, the $T_{s2}$ transition is not visible in the $C/T$ data. Probably, the associated peak is weak and broad so that it has been "absorbed" into the broad background from phonons; the background happens to have a broad hump at around 100 K near $T_{s2}$. It is also noted that there is a small peak at 250 K, which seems to correspond to the small anomaly in the $\chi$ data of Fig. 5b: there is a possibility of another phase transition at 243 K, although it is not conclusive.

The heat capacity at low temperatures is shown in a $C/T$ vs $T^2$ plot in the inset of Fig. 6. Apparently, there is a finite $C/T$ value when the curve is extrapolated to zero temperature. As usually assumed, we fit the data between 2 and 5 K to the equation $C = \gamma T + \beta T^3 + \alpha T^5$, where $\gamma$ is the Sommerfeld coefficient. The thus-obtained $\gamma$ is 8.3(4) mJ K$^{-2}$ mol$^{-1}$, which is larger than those of simple metals, smaller than 30 mJ K$^{-2}$ mol$^{-1}$ for CRO, and much smaller than those of strongly correlated electron systems. The Debye temperature $\Theta_D$ is estimated to be 288 K from $\beta$ = 0.730(7) mJ K$^{-4}$ mol$^{-1}$, which is a typical value for transition metal oxides.

One measure for the magnitude of electron correlations $U$ is the Wilson ratio $R_W$, which is calculated as $R_W$ = 72.95($\chi_0/\gamma$), where $\chi_0$ is $\chi$ at $T$ = 0 [18]. Since both $\chi_0$ and $\gamma$ are proportional to the DOS and only $\chi_0$ is enhanced by electron correlations, $R_W$ increases from 1 for $U$ = 0 to 2 for the large limit of $U$. Provided that $\chi_0$ is approximately 1.0 × 10$^{-4}$ cm$^3$ mol$^{-1}$ from the 2 K value, $R_W$ of phase III of PRO is 0.9. This small value indicates that electron correlations play a minor role in PRO, as is the case for CRO [5]. Related to this, we did not observe $T^2$ resistivity at low temperatures, which is also consistent with weak electron correlations. Note in Fig. 5b that $\chi$ increases gradually upon cooling below $T_{s2}$. The reason for this is not known but it may convey important information on the electronic states of the LT phases

### 3.2. Structural transitions

We have investigated the possibility of structural transitions at low temperatures by high-resolution powder XRD experiments. Figure 7 shows the temperature evolution of the XRD profile in a range of diffraction angles containing the two intense reflections of the (1 1 3) and (0 3 0) indices for the HT $R$–$3m$ structure. The two peaks move slightly to the high-angle side with decreasing temperature for 300–270 K, while, for each of them, a new peak appears and grows at a higher angle below 260 K. At 160 K, the profile is fitted to four peaks that are indexed as (2 2 2), (0 1 3), (3 1 2), and (4 0 –2) from the low-angle side assuming a monoclinic cell. Thus, a structural transition takes place at $T_{s1}$ from the $R$–$3m$ to a monoclinic structure; the possibility of a triclinic structure is not excluded.

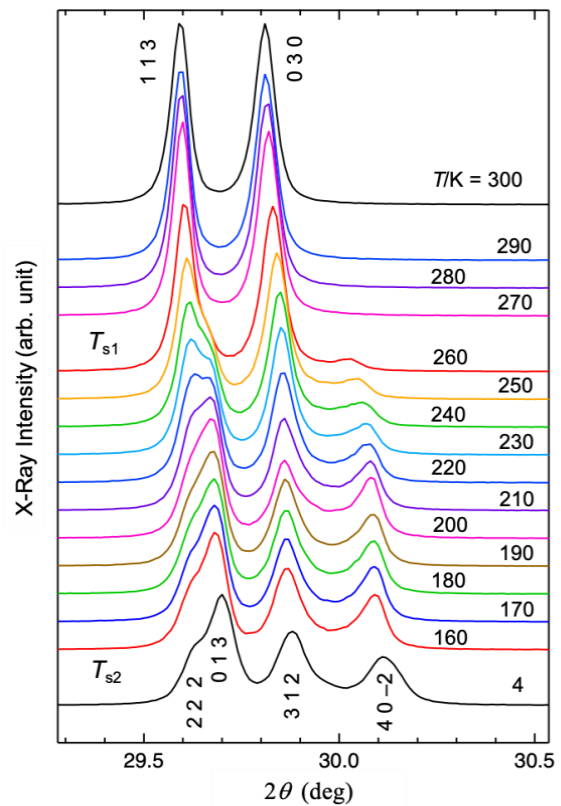

**Fig. 7.** Temperature evolution of the powder XRD profiles using the Cu–K$\alpha_1$ radiation from crushed crystals. The angle range includes the (1 1 3) and (0 3 0) reflections for the HT $R$–$3m$ structure. For clarity, each curve is shifted downward by appropriate offsets. The lowest-temperature profile at 4 K is reproduced by assuming a

monoclinic unit cell of $a$ = 13.9775(4) Å, $b$ = 10.3673(2) Å, $c$ = 9.4811(3) Å, $\beta$ = 96.7511(15)°.

Below $T_{s2}$, on the other hand, no further substantial changes were observed, which suggests that there is no structural transition at $T_{s2}$ or a subtle one to another monoclinic structure. The monoclinic lattice parameters at 4 K are $a$ = 13.9775(4) Å, $b$ = 10.3673(2) Å, $c$ = 9.4811(3) Å, $\beta$ = 96.7511(15)°. At this stage, we have not been successful in determining the space groups and obtaining the structural parameters for the LT phases. In particular, it is important to test whether the inversion symmetry is lost or not at the transitions. Experiments that are sensitive to the presence of inversion symmetry are now being undertaken, such as optical second-harmonic generation and convergent-beam electron diffraction experiments.

*3.3. Comparison to CRO*

We have shown that PbRe$_2$O$_6$ resembles Cd$_2$Re$_2$O$_7$ in various aspects: two successive transitions at similar temperatures, a large reduction in the DOS and a structural transition to low symmetry at $T_{s1}$, and relatively small changes at $T_{s2}$. One difference is that the $T_{s1}$ transition is of the first order in PRO, in contrast to the second-order transition in CRO; the $T_{s2}$ transitions are of the first-order in both compounds. It is remarkable that such similarities occur in spite of the substantial differences in the crystal structure and probably in the electronic structure, even though both are isoelectronic with the 5$d_2$ electron configuration. This means that a common Fermi liquid instability for the SOCM dominates their properties.

Concerning the structural transition at $T_{s1}$ in PRO, we think that it is not ascribed to a simple structural instability but an electronic instability, because the insulating analogue, PbNb$_2$O$_6$, shows a structural transition to a different polar structure of $R3m$ [19]. This is also the case for CRO: the insulating analogue Cd$_2$Nb$_2$O$_7$ exhibits a different transition scheme of $Fd$–$3m \rightarrow Ima2 \rightarrow Cc$ upon cooling [20], which is distinguished from the scheme of $Fd$–$3m \rightarrow I$–$4m2 \rightarrow I4_122$ for CRO. In contrast to these SOCM candidates, LiOsO$_3$ and its insulating analogue LiNbO$_3$ crystallize in the same $R3c$ structure [11]. Based on these comparisons between PRO and CRO, we consider that PRO is the second example of the SOCM. Future study would clarify this issue and uncover the general characteristics of SOCMs.

**4. Summary**

In summary, we have shown that PbRe$_2$O$_6$ can be a candidate SOCM, which comprises a stack of modulated honeycomb lattices made of the Re$^{5+}$ (5$d_2$) ions in the centrosymmetric $R$–$3m$ structure at room temperature. Two successive phase transitions are observed at $T_{s1}$ = 265 K and $T_{s2}$ = 123 K. At $T_{s1}$, the DOS is enormously reduced and a structural transition to a monoclinic structure takes place, while small changes are observed at $T_{s2}$. The similarity between PbRe$_2$O$_6$ and another SOCM candidate Cd$_2$Re$_2$O$_7$ suggests that there is a common Fermi liquid instability associated with SOCMs.

**Acknowledgment**

The authors are grateful to M. Suzuki for fruitful discussion. This work was partly supported by KAKENHI (Grant Number JP18K13491 and JP18H04308) and by Core-to-Core Program (A) Advanced Research Networks given by Japan Society for the Promotion of Science (JSPS).